\documentclass[a4paper,11pt]{article}
\usepackage{pos}
\usepackage{subcaption}
\usepackage{graphicx}
\usepackage{tikz}

\title{Multi-messenger Observations of Tidal Disruption Events}

\author*[a,b]{Simeon Reusch}

\affiliation[a]{Deutsche Elektronen-Synchrotron,\\
  Platanenallee 6, 15738 Zeuthen, Germany}

\affiliation[b]{Insitut für Physik, Humboldt-Universität zu Berlin,\\
Newtonstraße 15, 12489 Berlin, Germany}

\emailAdd{simeon.reusch@desy.de}

\abstract{Using the Zwicky Transient Facility (ZTF) and other observatories, we have identified three candidate Tidal Disruption Events (TDEs) in spatial and temporal coincidence with high-energy neutrinos detected by IceCube: AT2019dsg, AT2019fdr and AT2019aalc. All three of these events have been shown to be able to produce high-energy neutrinos. In these proceedings, I will give an overview of Tidal Disruption Events, outline our follow-up program with ZTF, describe the observations carried out for each of those coincident events and highlight their similarities and differences.}

\FullConference{%
   27th European Cosmic Ray Symposium - ECRS \\
   25-29 July 2022 \\
   Nijmegen, the Netherlands 
}

\begin{document}

\maketitle
\section{Tidal Disruption Events}
Tidal Disruption Events (TDEs) have already been predicted in the 1970s \cite{hills}. The underlying idea could come from a curious child's mind: What happens if a star orbiting the supermassive black hole (SMBH) at the center of a galaxy falls into the black hole? At first glance, the answer is straightforward: When the orbiting star gets close to the black hole, the resulting tidal forces from the black hole grow larger than the star's self-gravity and destroy it \cite{rees_tde}. Roughly half of the star's mass gets accreted around the black hole and the bright electromagnetic flare resulting from this accretion can last for months. However, the exact mechanisms at play which cause this emission are still broadly discussed.

\begin{figure}[hb]
    \centering
    \includegraphics[width=0.4\linewidth]{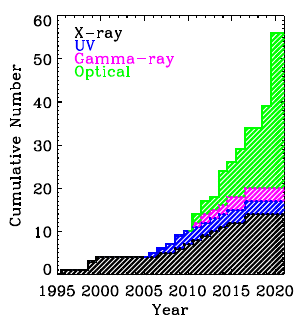}
    \caption{Cumulative TDE detection distribution, sorted by discovery wavelength. The relative increase of optical discoveries in 2020 is driven by ZTF. From \cite{gezari}.}
    \label{fig:tde_cumulative}
\end{figure}

To understand them, we need observational data. The first few TDEs have been discovered in the 1990s and early 2000s in X-ray wavelengths. For a long time, none were detected in optical searches. Only in the last decade that suddenly changed \cite{first_optical}, thanks to the advent of optical all-sky surveys such as Pan-STARRS \cite{panstarrs_tde}, ASASSN \cite{asassn_tde}, Gaia \cite{gaia_tde} or the Zwicky Transient Facility (ZTF) \cite{final_season}. The rising number of optically discovered TDEs can be seen in Fig. \ref{fig:tde_cumulative}. The large field of view and high cadence of these survey telescopes makes them the ideal discovery tools for TDEs. As of December 2022, roughly 100 TDEs have been discovered in total, with more than 30 detected during the 2.6 years of ZTF Phase I alone \cite{final_season}.

\begin{figure}
    \centering
    \includegraphics[width=0.8\columnwidth]{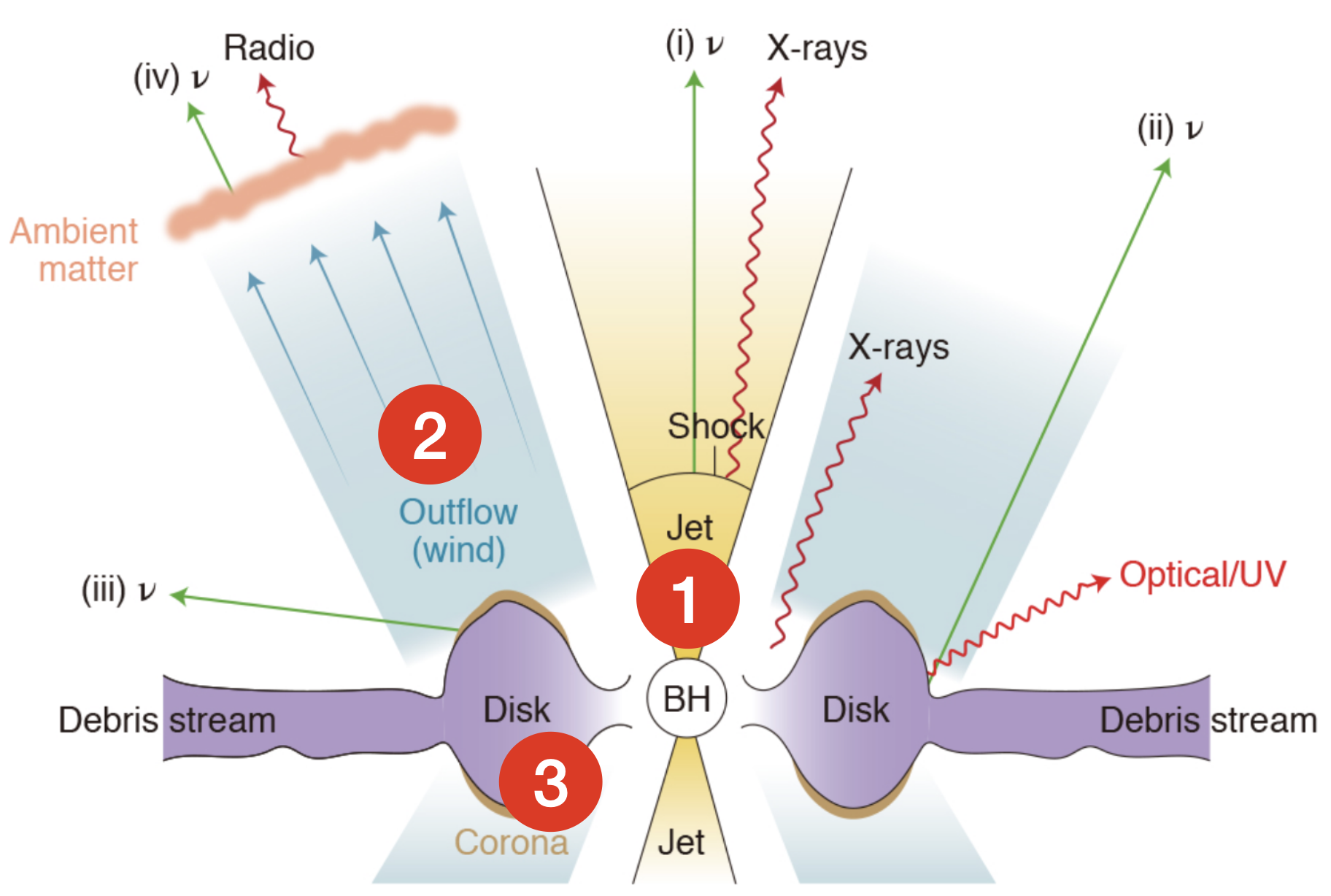}
    \caption{Unified TDE model (peak wavelength is dependent on viewing angle) as well as possible production sites for high-energy neutrinos: (1) relativistic or sub-relativistic jet, (2) semi-relativistic outflow, (3) corona of the newly formed accretion disk. From \cite{hayasaki}, annotations by author. Note that most TDEs are thought \textit{not} to produce jets.}
    \label{fig:unified_model}
\end{figure}

The observed continuum emission of TDEs can be approximated by a blackbody fairly well. However, the temperature distribution seems to be bimodal, with one population peaking in optical/ultraviolet (UV) wavelengths, and a second X-ray bright population best described by blackbodies with higher temperatures (see \cite{gezari} for a review). As the newly formed accretion disk from a TDE will most likely be very hot, it is not clear where the optical/UV emission comes from. Also, the blackbody radii inferred by this emission exceed the newly formed accretion disk's size by 1-2 orders of magnitude. So far, multiple models have been proposed to explain the optical/UV light, such as semi-relativistic outflows or winds, as well as shock acceleration stemming from the tidal stream intersecting itself. If the debris resulting from the tidal disruption is rapidly circularized, the bimodal distribution (X-ray vs. optical/UV) might be reconciled by a unified TDE model, as proposed in \cite{unified_model}. In this model, the most prominent wavelength detected depends on the viewing angle, as shown in figure \ref{fig:unified_model}: When one looks into the funnel perpendicular to the disk, one can see the X-rays from the inner disk. From a side-on view, the X-rays will be obscured and one can only see the optical and UV emission stemming from X-rays reprocessed in the outer disk or in outflows. Intermediate viewing angles will produce a mixture of both signals.

Additionally, around 1\% of TDEs launch relativistic jets, like e.g. the recently discovered AT2022cmc \cite{jetted_tde}. Such jets (labeled as (1) in Fig. \ref{fig:unified_model}), as well as the possible winds or outflows (2), or a potentially present disk corona (3), have been proposed as production sites for high-energy neutrinos, making TDEs promising candidates for multi-messenger emission. In all of these scenarios, protons are efficiently accelerated to high energies and subsequently interact with other protons or a photon target field ($pp$ or $p\gamma$ interaction) \cite{hayasaki}.

\section{Our search for sources of high-energy neutrinos}
Detecting neutrinos is notoriously hard, as they only interact via the weak nuclear force. In most cases, they are detected indirectly using large reservoirs of matter. In the case of the IceCube detector, the medium of choice is (surprise) ice. Charged-current interactions of muon neutrinos with the ice produce muons. These muons emit Cherenkov radiation while travelling through the ice, as their speed exceeds the speed of light in this medium. This light is then detected by photomultiplier tubes within the ice. Using the light intensity and timing information, one can reconstruct the direction of origin of the neutrino. The reconstructions of these muon tracks typically result in 90\% rectangular uncertainty areas ranging from a few to a few tens of square degrees. Other neutrino interactions produce spherical light patterns in the detector, which do not allow for good angular reconstructions.

Recently, the nearby galaxy NGC1068 made the headlines because of a correlation with the full set of astrophysical neutrinos detected by IceCube with an excess of 79 neutrinos with TeV energies at a $4.2 \sigma$ level \cite{ngc_1068}. It has also been possible to identify counterparts to single high-energy neutrino alerts.
So far, though, only a handful of high-energy neutrinos have been tied to likely sources. The most prominent one was IC170922A, for which the flaring blazar TXS 0506+056 was identified as the likely origin \cite{txs_mm}.

To find more candidate counterparts, ZTF has been operating a systematic optical follow-up program since 2019, targeting the 90\% sky localizations of selected high-energy neutrino alerts from IceCube. ZTF is an optical survey telescope located at Mt. Palomar in California, with an average depth of 20.5 mag. It has a very large field of view of 47 sq. deg, dwarfing all other optical survey telescopes, including the future Rubin Observatory. This makes it a great choice for follow-up, as usually one pointing per band is enough to cover the respective uncertainty areas \cite{ztf_science_objectives}. We receive the alerts from IceCube via the low-latency Gamma-ray Coordination Network (GCN). When either the probability of the neutrino of not being atmospheric background is larger than 50\% or its uncertainty area is smaller than 10 sq. deg, we follow up the alert, provided the sky location is accessible to ZTF and the uncertainty area is smaller than 40 sq. deg. We react as fast as possible, with 300 second observations (with a typical depth of 21.5 mag) in the first night, followed by shallower 30 second observations during subsequent nights to monitor the time evolution of our candidates. So far, 30\% of all alerts have led to observations, with a total number of 32 observational follow-up campaigns as of December 2022.

\begin{figure}
    \centering
    \includegraphics[width=0.8\columnwidth]{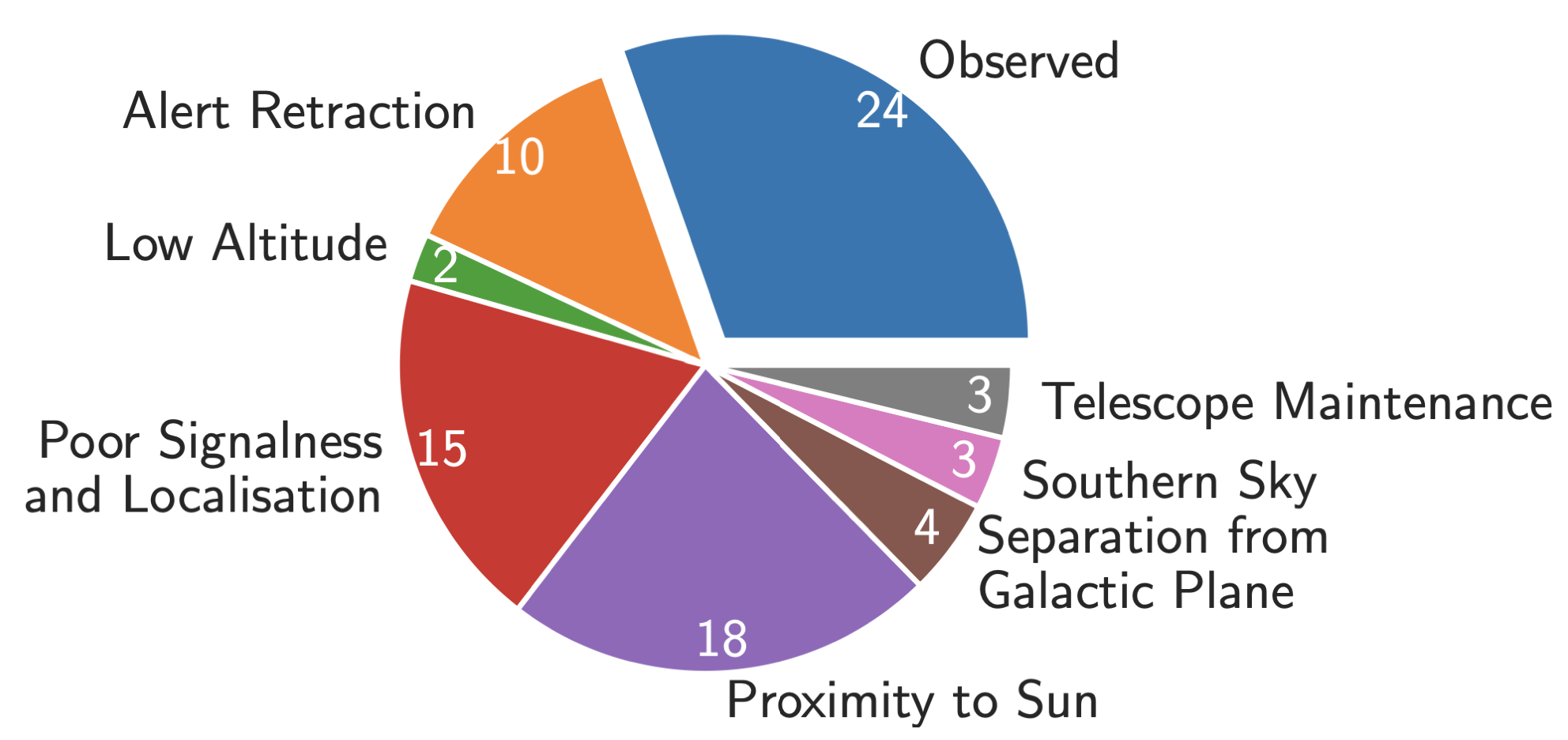}
    \caption{Follow-up performance as of 21 Dec 2021. \textit{Poor Signalness and Localisation} refers to the alert quality cut described below. From \cite{fu_program}.}
    \label{fig:fu_performance}
\end{figure}

The transients we get from observing are then filtered with \texttt{nuztf} \cite{nuztf}, our dedicated follow-up pipeline. This pipeline makes heavy use of \texttt{AMPEL}, a broker and analysis framework designed to facilitate the live and archival processing of large numbers of astronomical transients \cite{ampel}. Considering the vast number of transient alerts generated by ZTF each night, automated filtering is strictly necessary. We cut down the large number of alerts by requiring that candidates a) lie within the 90\% localization region of IceCube, b) are detected at least once \textit{after} the neutrino has been detected, c) have a sufficiently high probability of being real (not a subtraction artefact) and finally d) are likely not a solar system object or star. To increase the quality of the lightcurves and to be sensitive to pre-neutrino activity, we usually obtain forced photometry for candidate counterparts with \texttt{fpbot} \cite{fpbot}, our forced photometry pipeline.

The follow-up program is designed to be sensitive to the following high-energy neutrino production environments: Supernovae with evidence of interaction with their circumstellar medium, supernovae with relativistic jets, gamma-ray bursts, active galactic nucleus (AGN) activity that is correlated to optical flares, and TDEs. After observing and automated filtering, we vet the remaining candidates and trigger subsequent observations, most importantly spectroscopy, to classify them. For more details on the program, see \cite{fu_program}.

The remains of these proceedings will focus on the two (candidate) TDEs we found in temporal and spatial coincidence with high-energy neutrinos during the operation so far (AT2019dsg and AT2019fdr), as well as a third candidate that emerged during a subsequent analysis (AT2019aalc).

\section{Three candidate tidal disruption events associated with high-energy neutrinos}

\subsection{First association: AT2019dsg}
On October 1, 2019, we triggered follow up to IceCube neutrino IC191001A. Among the candidates selected by our pipeline was the several months old TDE AT2019dsg. As is the case for all ZTF detected TDEs, observations by NASA's \textit{Neil Gehrels Swift Observatory} \cite{swift} had already been carried out with the Ultra-Violet/Optical Telescope (UVOT) \cite{swift_uvot}, revealing bright emission in the UV tracing the brightness evolution of the optical wavelengths in the lightcurve. The blackbody temperature inferred from the optical/UV measurements was $10^{4.6}~\rm K$, somewhat hot for a TDE \cite{at2019dsg}.

Additionally, observations had been carried out with the X-ray Telescope (XRT) \cite{swift_xrt} aboard \textit{Swift}, showing soft X-ray emission fading to non-detection within 70 days after discovery. This could either be explained by cooling of the accretion disk or by an obscuration effect. The X-ray detection is not uncommon (9 of the 30 ZTF Phase-I TDEs are detected in X-rays) \cite{final_season}. A more peculiar feature is the radio detection of AT2019dsg. Long-lasting radio emission has been detected by a variety of instruments: The Large Array of the Arcminute MicroKelvin Imager (AMI) \cite{ami1, ami2}, MeerKAT \cite{meerkat} and the Karl G. Jansky Very Large Array (VLA) \cite{vla}. These showed temporal evolution over the course of months after the detection of the neutrino. The interpretation of the radio signal has been somewhat disputed, but at the very least it confirms long-lived non-thermal emission \cite{at2019dsg}.

Lastly, later analysis of data from the \textit{Wide-field Infrared Survey Explorer} (WISE) \cite{wright2010} showed a prominent infrared (IR) flux increase with respect to epochs prior to the TDE -- a feature I will come back to.

\subsection{Second association: AT2019fdr}
Half a year later, AT2019fdr was found in coincidence with IceCube neutrino IC200530A. At the time of neutrino detection it was already 10 months old, and its nature somewhat disputed. Within the ZTF collaboration, it was initially classified as a superluminous supernova of type II (SLSN II), though a study of peculiar accretion flares in Narrow-line Seyfert Galaxies (NLSy1) argued for a TDE nature of the flare on the basis of AT2019fdr's long-lived and bright U-band and UV emission, the proximity to the core of its host galaxy as well as emission at the blue end of the Balmer line profiles and the flare's overall longevity \cite{sara}. To this body of evidence we added a late-time X-ray detection and a bright infrared dust echo, which makes a SLSN II interpretation even less likely (see below).

\begin{figure}
    \centering
    \includegraphics[width=1\columnwidth]{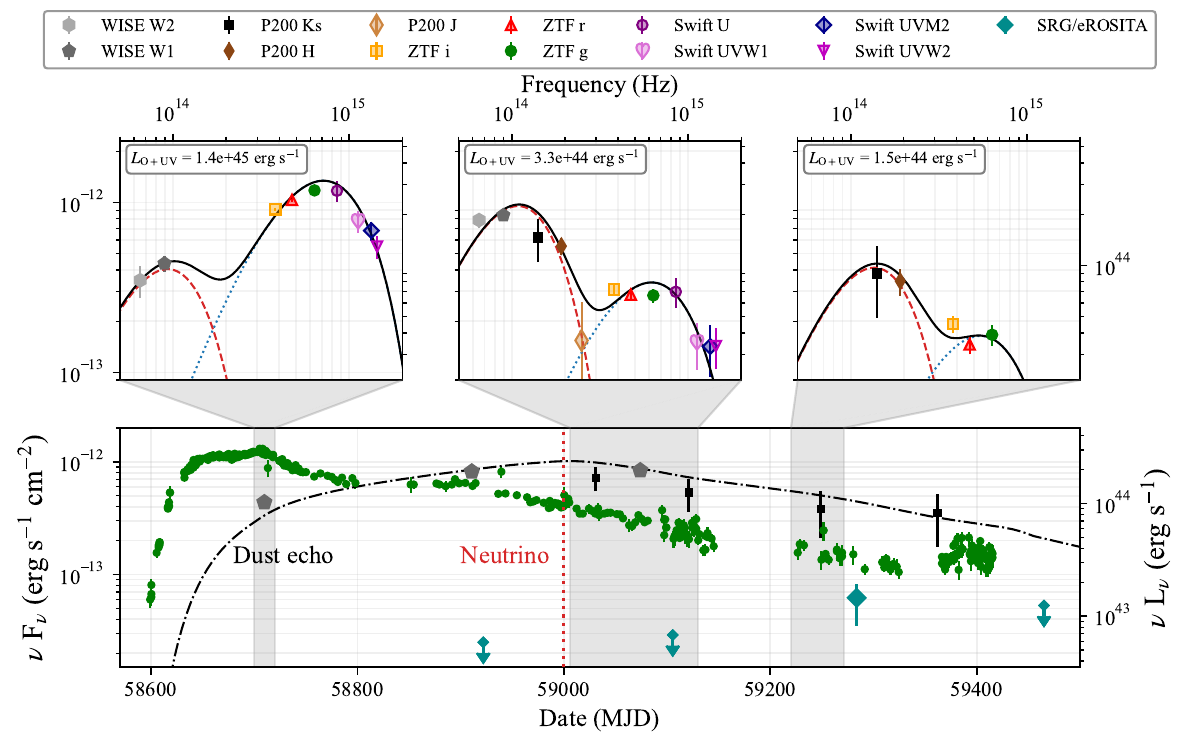}
    \caption{SEDs and lightcurve of AT2019fdr. The three top panels show the SED at different times as well as the fitted NIR and optical/UV blackbodies. The neutrino arrival time is shown as red vertical dotted line in the lightcurve. The inferred dust echo lightcurve is shown as black dash-dotted line. Luminosities are given in the source rest frame. From \cite{at2019fdr}.}
    \label{fig:moneyplot}
\end{figure}

We think that AT2019fdr can be classified as TDE, albeit an unusual one. Its peculiarity should not be surprising, as it is located in an AGN, a somewhat different environment than a quiescent host galaxy. AT2019fdr's lightcurve and spectral energy distributions (SED) during three different epochs can be seen in Fig. \ref{fig:moneyplot}. Each the optical/UV and the infrared part of the SED are well approximated by a blackbody. Note that AT2019fdr is one of the most luminous transients ever discovered: From the optical/UV blackbody one can infer a total bolometric energy of at least $10^{52}~\rm erg$ (after fitting for extinction). Fig. \ref{fig:moneyplot} also shows a late-time detection by eROSITA \cite{erosita} aboard \textit{Spektrum Roentgen-Gamma} (SRG) \cite{erosita_xrg}. This detection displayed a very soft thermal spectrum with a blackbody temperature of only $56^{+32}_{-26}~\rm eV$. As X-ray emission is very unusual for SLSNe, we count this as further evidence for AT2019fdr being a TDE.

\begin{figure}[htbp]
    \centering
        \includegraphics[clip, trim=12cm 7.6cm 12cm 5cm, width=0.8\textwidth]{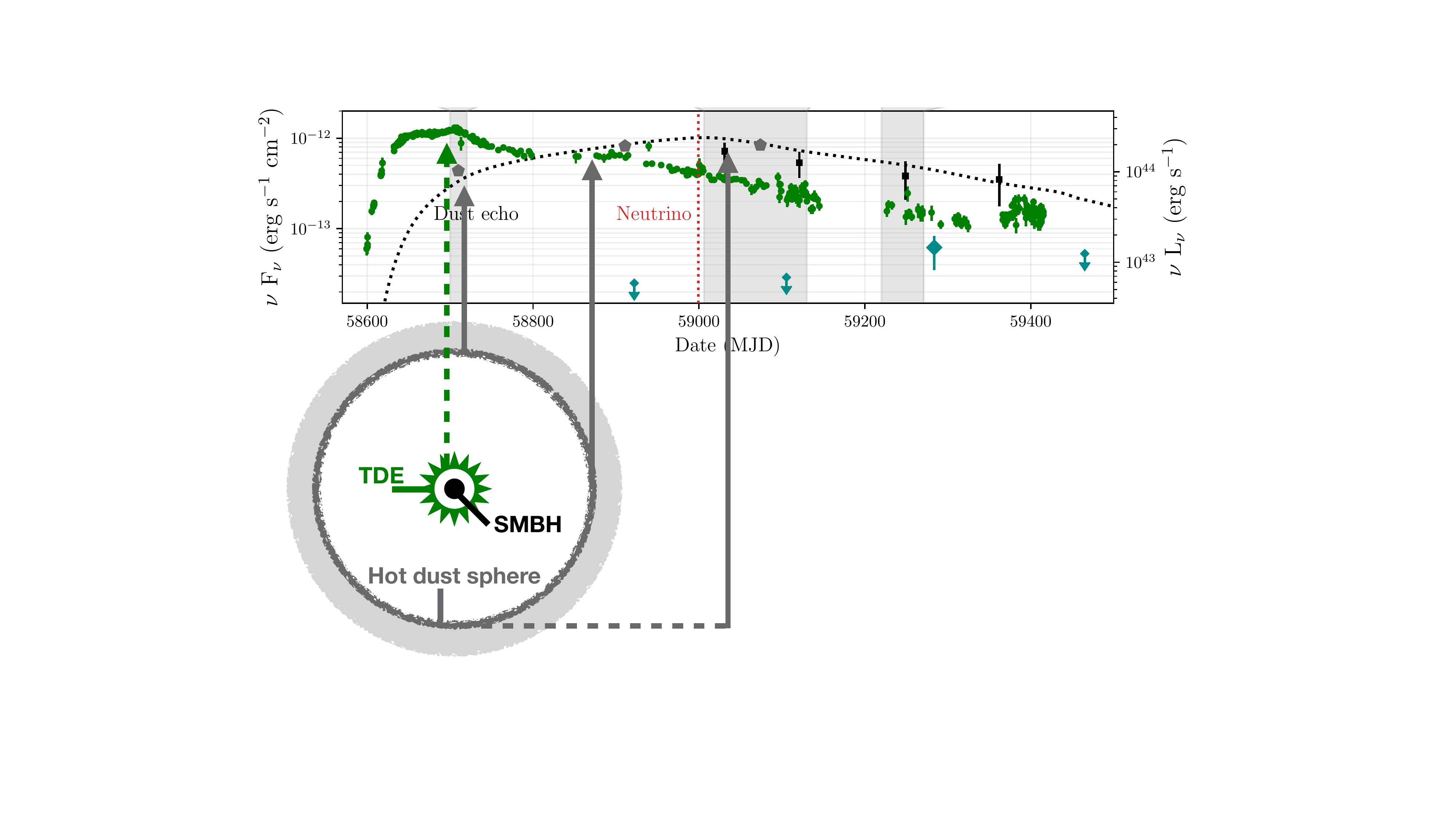}
    \caption{AT2019fdr: Illustration of the light travel time induced delay of the IR dust echo signal. Light from the TDE around the supermassive black hole (SMBH) gets reprocessed by a hot dust sphere, which emits thermally in the IR. IR light directly from the line of sight arrives first, and light from the back of the system arrives last, explaining the stretching of the IR lightcurve with respect to the optical one.}
    \label{fig:dust_system}
\end{figure}

We complemented the mid-infrared \textit{WISE} measurements with near-infrared observations with the Wide Field Infrared Camera (WIRC) \cite{wirc} mounted on the P200 telescope at Mt. Palomar. These comprise a blackbody, which together with the delayed evolution of the IR signal with respect to the optical flare strongly suggests that we see a dust echo. This is the reprocessing of the flare by a surrounding dust region that gets hot and begins to emit thermally after a delay caused by the light travel time: Light along the line of sight will arrive simultaneous to the optical signal, but light perpendicular to it will arrive later. Finally, light from the back of the system will arrive. A fitted dust echo and a sketch illustrating the dust system can be seen in Fig. \ref{fig:dust_system}. From this fit and by comparing the optical/UV and IR luminosities, one can infer the distance of the dust shell to the core ($0.16~\rm pc$) and a covering factor of $1/3$. This value is rather large compared to TDEs in quiescent galaxies, but in good agreement with comparable events \cite{hinkle_dust}. The discovery of this dust echo, alongside the large energy budget and slow temporal evolution, suggests that AT2019fdr belongs to a class of TDE candidates occurring in AGN; an especially violent environment which readily explains the amounts of dust required for such a luminous dust echo with a high covering factor. It can also be counted as evidence for the TDE interpretation, as the amount of dust intrinsically produced by SNe is fairly limited, which is hard to reconcile with the covering factor observed for AT2019fdr \cite{ps1-adi-echo}.

Furthermore, radio measurements obtained with VLA over multiple epochs showed a consistent signal. However, as it did not display signs of time evolution, the radio emission most likely stems from the underlying AGN.

The combined chance coincidence for finding AT2019dsg and AT2019fdr in coincidence with high-energy neutrino alerts is 0.03\%. Note that a stacking analysis from 2019 has constrained the contribution of TDEs to the observed diffuse IceCube neutrino flux to $\leq 30\%$ \cite{robert_2019}. If one accounts for TDEs and candidate TDEs like AT2019fdr, at least 7.8\% of astrophysical IceCube high-energy neutrinos would come from this broader population \cite{at2019fdr}.

\subsection{Third association (from dust echo study): AT2019aalc}
The discovery of two prominent dust echoes prompted a search for other optical flares accompanied by a \textit{WISE}-detected dust echo in coincidence with high-energy neutrinos. A third candidate counterpart emerged from this search, which was previously missed as its location was not observable by ZTF at the time of neutrino detection: AT2019aalc, probable counterpart to neutrino IC191119A.

Much less is known about this event as it flew under the radar for quite a while. It was discarded as probable AGN activity during scanning and was not scrutinized further until it emerged as neutrino counterpart candidate months later. In AT2019aalc's case, the neutrino arrived about 5 months after the optical peak. The lightcurve is consistent with this event also being a TDE, though this classification is much less secure compared to AT2019fdr, as there are no spectra or UV measurements available. The peak g-band luminosity of AT2019aalc is comparable to AT2019dsg, and its dust echo luminosity is the largest of all three events -- although its dust echo strength as defined in \cite{at2019aalc} is lower due to the pre-flare variability. Interestingly, it was also detected by SRG/eROSITA on a regular visit, displaying a fairly soft spectrum with a blackbody temperature of $172\pm10~\rm eV$. This can be counted as evidence in favor of a TDE interpretation. There was also an archival radio detection by the VLA Sky-Survey (VLASS) \cite{vlass}.

The significance of finding three events in spatial and temporal coincidence with a high-energy neutrino is $3.7\sigma$ \cite{at2019aalc}. Note that the soft X-ray emission detected for all transients was not part of the initial selection criteria.

\section{Comparison}
A comparison of the relevant measurements and inferred quantities for the three events is shown in Table \ref{tab:comparison}. All three events have in common that the neutrino was emitted significantly later than the optical peak (with a delay of 5-10 months). Furthermore, all were detected in X-rays with fairly low temperature, and all showed prominent dust echoes. One can try to draw conclusions for production scenarios from these facts.

\begin{figure}[htb]
\centering
\begin{subfigure}{0.5\textwidth}
  \centering
  \includegraphics[width=1.1\linewidth]{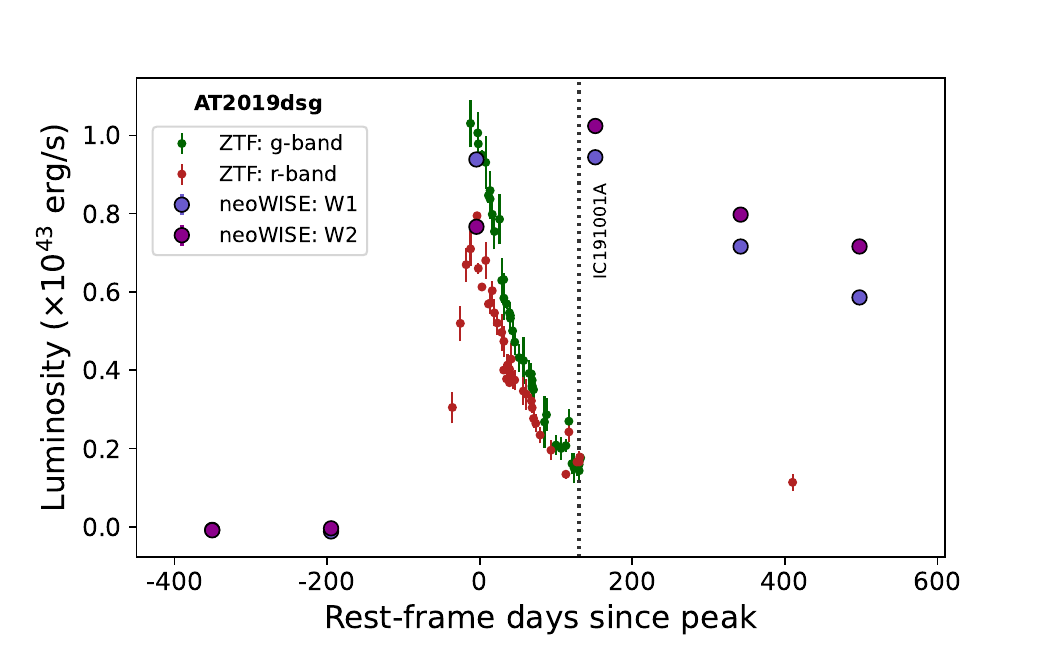}
  \caption{AT2019dsg}
  \label{fig:echo_1}
\end{subfigure}%
\begin{subfigure}{0.5\textwidth}
  \centering
  \includegraphics[width=1.1\linewidth]{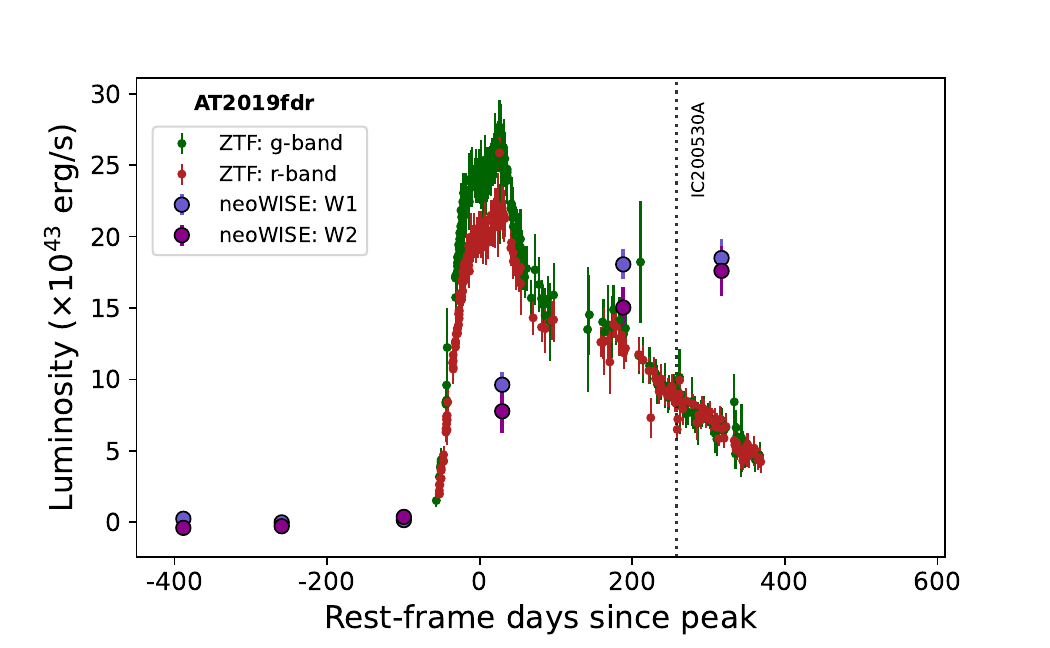}
  \caption{AT2019fdr}
  \label{fig:echo_2}
\end{subfigure}
\begin{subfigure}{0.5\textwidth}
  \centering
  \includegraphics[width=1.1\linewidth]{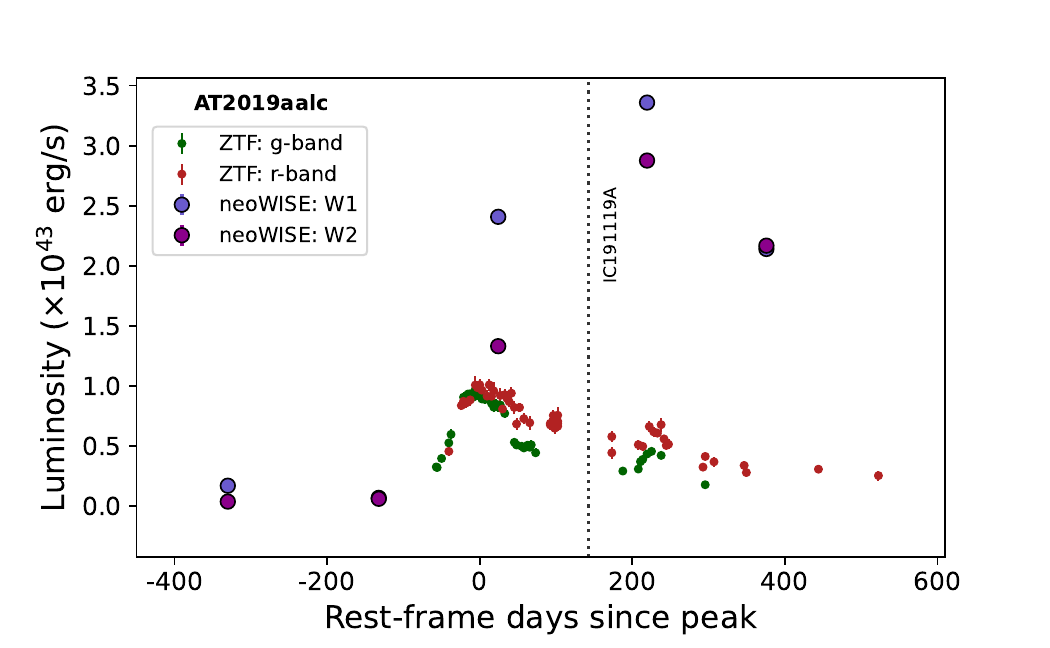}
  \caption{AT2019aalc}
  \label{fig:echo_3}
\end{subfigure}
\caption{All three candidates with optical lightcurve and \textit{WISE} detections, showing the relative strength of the dust echo. The neutrino arrival times are marked by dotted vertical lines. Figures adapted from \cite{at2019aalc}.}
\label{fig:echo}
\end{figure}

The observed delay might either be a purely statistical effect, but could also carry physical meaning. In \cite{at2019aalc} this is explained by assuming that the debris first needs to circularize before the first neutrinos can be produced. In \cite{walter_dust}, the observed neutrino delay is combined with the dust echo present in all three candidate counterparts and which is peaking around the time of neutrino detection in all three cases (see Fig. \ref{fig:echo}). In one of the models presented, the infrared dust-echo photons serve as the target for accelerated protons. In this scenario the neutrino delay arises naturally from the delayed IR emission. The energies involved would render TDEs interesting candidates for Ultra-High Energy Cosmic Ray (UHECR) emission. This model comes at the cost of requiring very high proton energies. A companion model proposing X-ray target photons, which are also available for all three sources, can explain the neutrino delay too. In this case, the delay arises from the confinement of protons with moderate energy. This model explains the observed neutrino energies better, at the cost of describing the observed neutrino delay less good than the IR model. A third model uses the optical/UV emission as target (similar to \cite{at2019dsg}). This has the advantage of yielding the highest neutrino production efficiency of all three models, but fails to describe the observed neutrino time delay.

To decide if one of these models is correctly describing the possible neutrino production mechanisms in TDEs will require more candidate counterparts to test the predictions \cite{walter_dust}.

\begin{table}[htb]
\centering
\setlength{\tabcolsep}{12pt}
\begin{tabular}{c  c  c  c} 
\textbf{Property} & \textbf{AT2019dsg} &\textbf{AT2019fdr} & \textbf{AT2019aalc}\\
\hline
TDE & yes & strong candidate & candidate \\
Peak bol. luminosity & $3.5 \times 10^{44} \rm erg~s^{-1}$ & $1.3 \times 10^{45} \rm erg~s^{-1}$ & -- \\
SMBH Mass & $10^{6}-10^{6.7} M_\odot$ & $10^{7.55} M_\odot$ & $10^{7.2} M_\odot$ \\
Radio & evolving & not evolving & archival det. \\
UV & very bright & bright & -- \\
X-ray & early, soft spectrum & late, soft spectrum & soft spectrum \\
Dust echo strength & $92.2$ & $39.2$ & $15.7$ \\
$\nu$ delay & ~5 months & ~10 months & ~5 months \\
$\nu$ production & possible & possible & possible \\
$\nu$ energy & 217 TeV & 82 TeV & 176 TeV \\
$\nu$ 90\% uncertainty box & 25.5 sq. deg & 25.2 sq. deg & 61.2 sq. deg \\
$\nu$ signalness & 0.59 & 0.59 & 0.45 \\
\hline
\end{tabular}
\caption{Comparison of the multi-messenger properties of the three candidate TDEs detected in coincidence with high-energy neutrinos. Dust echo strength is defined as $\Delta F/F_{\rm RMS}$ (flux increase after optical/UV peak vs. pre-flare root mean square); \textit{$\nu$ production} means that models exist that show that the source is capable of producing a neutrinos of the respective detected energy; $\nu$ signalness is the probability of the neutrino of being of astrophysical origin (as opposed to atmospheric). -- denotes insufficient modeling or missing data. Numbers from \cite{at2019dsg, at2019fdr, at2019aalc}; possibility of neutrino production for AT2019aalc from \cite{walter_dust}.}
\label{tab:comparison}
\end{table}

\section{Summary and Outlook}
So far, three accretion flares have been found in temporal and spatial coincidence with high-energy neutrinos detected by IceCube. One of them was a bona fide TDE (AT2019dsg) and two were candidate TDEs (AT2019fdr and AT2019aalc). In all three cases, the detected neutrino was delayed with respect to the optical peak and an X-ray signal with a soft spectrum was detected. All three optical flares were accompanied by a large infrared echo pointing at significant amounts of dust surrounding the black hole.

How can we be sure that these associations between the neutrinos and the events hold? The answer is easy: Continue the systematic follow-up program, and see what happens. If we can make more associations, the chance coincidences will keep dropping. If nothing new turns up, then that is just the way the universe works. AT2019fdr is interesting nevertheless, as its sheer brightness makes it highly unusual. This event is also remarkably long-lived, as we are still detecting it as of December 2022, over three years after the optical peak. Furthermore, it recently showed signs of optical rebrightening. There is still a lot to be learned about TDEs in AGN.
\newpage
\bibliographystyle{JHEP}
\bibliography{main}

\end{document}